# A Novel Method for Clustering Cellular Data to Improve Classification

Diek W. Wheeler and Giorgio A. Ascoli*

Center for Neural Informatics, Structures, & Plasticity, Krasnow Institute for Advanced Study; and Bioengineering Department, Volgenau School of Engineering; George Mason University, Fairfax, VA 22030-4444, USA

*For correspondence: ascoli@gmu.edu

**Abstract**

Many fields, such as neuroscience, are experiencing the vast proliferation of cellular data, underscoring the need for organizing and interpreting large datasets. A popular approach partitions data into manageable subsets via hierarchical clustering, but objective methods to determine the appropriate classification granularity are missing. We recently introduced a technique to systematically identify when to stop subdividing clusters based on the fundamental principle that cells must differ more between than within clusters. Here we present the corresponding protocol to classify cellular datasets by combining data-driven unsupervised hierarchical clustering with statistical testing. These general-purpose functions are applicable to any cellular dataset that can be organized as two-dimensional matrices of numerical values, including molecular, physiological, and anatomical datasets. We demonstrate the protocol using cellular data from the Janelia MouseLight project to characterize morphological aspects of neurons.

**Introduction**

Biomedical data in many fields are accumulating at ever increasing rates[1–3], and neuroscience is no exception[4,5]. In particular, large numbers of individual cells are now routinely characterized with high-throughput molecular sequencing and microscopic imaging, prompting multiple large-scale projects for comprehensive cellular classification across organisms and biological systems[6,7]. The most popular approach to managing a cumbersome dataset is to partition it into more intelligible parts using hierarchical clustering[8–12]. However, there is no agreed-upon objective method to determine the appropriate granularity of clusters[13]. Instead, after generating a clustering dendrogram from a dataset, an expert, but somewhat arbitrary, scientific judgement typically guides the final data classification by drawing a single horizontal line across the dendrogram[14,15]. We recently devised an original technique to improve this practice based on the fundamental principle that cells must differ more significantly between than within classes[16]. This simple reasoning can be implemented algorithmically by combining unsupervised hierarchical clustering with rigorous statistical test of variance[17], yielding an intrinsically data-driven classification.

The main protocol describes the execution of this enhanced cellular classification starting from any dataset organized as a two-dimensional matrix of numerical values, with rows and columns respectively representing individual cells and their features, such as gene expression or morphological traits. As customary, pairwise distances between the horizontal vectors of cellular values undergo unsupervised hierarchical clustering to generate a dendrogram (Figure 1). Starting from the top of the dendrogram (encompassing all cells) and proceeding down, at every branching point the (remaining) data are shuffled to generate an equivalent but homogeneous distribution of values. Next, the distribution of pairwise vector distances from the original data is compared to the distribution of pairwise vector distances from the shuffled data. The rationale is that, if the original data consists of (at least) two classes, its pairwise distances will include both (smaller) intra-class differences and (larger) inter-class differences, thus yielding a wider distribution than that resulting from the shuffled data, which instead represent the "one-class" null hypothesis. Specifically, Levene's one-tailed statistical test rigorously assesses whether the variance of the original distribution is significantly greater than the variance of the shuffled distribution. If so, the original data are partitioned into two separate classes in accordance with the clustering dendrogram, and the procedure repeats to analyze each of the distinct branches independently. If not, progress ceases down this branch of the dendrogram, and all cells below this point are classified as belonging to the same cluster. Note that this enhanced classification scheme creates a multi-tiered partitioning of the clustering dendrogram, because each branch will stop at its own distinct, data-driven granularity, thus providing a more refined data classification.

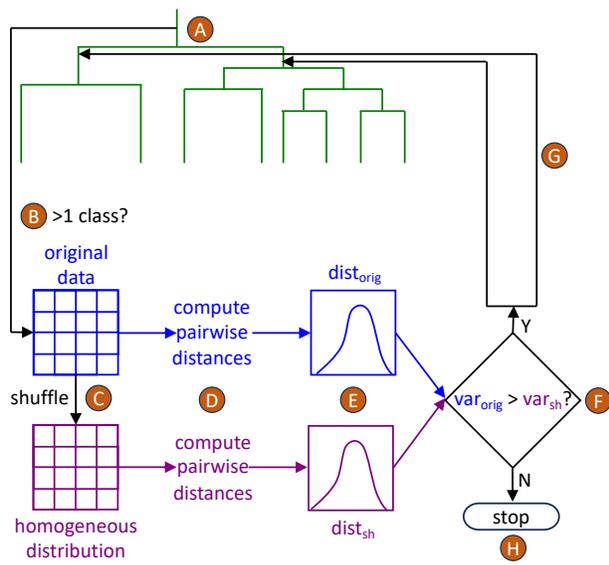

Figure 1. Pipeline for combining unsupervised hierarchical clustering with Levene's one-tailed statistical test to reveal the inherent classifications in cellular data. A) The set of pairwise distances between the horizontal vectors of the two-dimensional data matrix undergoes unsupervised hierarchical clustering to generate a dendrogram. B) The first branching point in the dendrogram is evaluated: can the cellular data be partitioned into at least two classes? C) The matrix of original data is shuffled to generate an equivalent distribution of homogeneous data values, corresponding to the null (one-class) hypothesis. D) The pairwise vector distances from the original and shuffled datasets are computed for statistical comparison. E) The widths of the original and shuffled distance distributions are quantified in terms of their variances. F) Levene's one-tailed statistical test is used to compare the original and shuffled variances. G) If the variance of the original distribution is statistically greater, then the data are partitioned into two separate classes according to the clustering dendrogram, and the process is repeated at each of the subsequent branches. H) If the original variance is not statistically greater, the procedure stops, and all cells located below the current branch are classified as belonging to the same cluster.

The secondary protocol presented here describes the execution of other functions of the same computer program, which were developed to analyze specific characteristics of the cellular data generated by the Janelia MouseLight project[18]. Based on the three-dimensional reconstructions of neurons, which are downloadable either from MouseLight (https://ml-neuronbrowser.janelia.org) or from the repository NeuroMorpho.Org[19], these data are numerical descriptions of how the parts of neurons, such as axonal and dendritic branches, are distributed across regions of the brain. The functions analyze the divergence and convergence of axonal projections, the soma distributions, the population sizes of identified clusters, and the Strahler order numbers.

Figures accompany both protocols to convey what the user can expect when executing the various functions of the program. The example data matrix referenced in these figures is from the Janelia MouseLight project and represents the amount of axonal presence the neurons have in each of the arrayed brain regions, which forms the basis of an approach for classifying neurons[20,21]. The program is designed to be executable on a basic laptop or desktop computer loaded with readily available software.

**Equipment.**

1. Standard personal computer: the described analyses were run on a 2019 16" MacBook Pro running macOS Ventura 13.6.1.

**Software.**

1. Custom-written MATLAB algorithms, in conjunction with exemplary data files, are freely accessible[22] through a GitHub repository (https://github.com/Projectomics/MATLAB).
2. MATLAB commercial license: the described algorithms were run with MATLAB 23.2.0.2365128 (R2023b).

**Image dataset and additional requirements.**

1. Janelia MouseLight reconstructions (https://www.janelia.org/project-team/mouselight/neuronbrowser/), both JSON and SWC versions[23].
2. Tract values from a source, such as the regional connectivity from anterograde tracing to the known targets, as presented in the Allen Mouse Brain Connectivity Atlas (http://connectivity.brain-map.org/projection).

**Protocol Formatting.**

1. File and directory names are in "double quotes."
2. MATLAB commands or functions are in *italics*.
3. Variable names are in the font Courier New.
4. Menu selections for the analyses are in 'single quotes.'
5. A labeled step in the protocol is in **bold font**.

**Procedures for General Modality-free Data.**

1. Organize the main input matrix in the manner of "PRE_matrix__axonal_counts_per_neuron_per_parcel.xlsx," where the number of counts for each neuron per category is stored and where the counts for each neuron compose a row vector whose components are the counts in each category. For example, neuron information is listed across each row, and parcel information listed down each column (Figure 2).

| Neurons \ Parcels | Presubiculum | Subiculum | fiber tracts | dorsal hippocampal commissure |
|---|---|---|---|---|
| AA0021 | 13 | 150 | 26 | 75 |
| AA0024 | 99 | 5 | 1 | 109 |
| AA0026 | 124 | 0 | 0 | 123 |
| AA0030 | 136 | 15 | 0 | 155 |
| AA0031 | 981 | 256 | 11 | 2 |

Figure 2. Main source data file.

2. Run *main_menu_selector()*. Including no variable arguments will default to setting the variables `nShuffles` = 100,000 and `isUseOriginalAlgorithm` = 1.
    i. `nShuffles` is the number of times the original data file is shuffled to generate a randomized version for the determination of the statistical significance of the value distributions in the original data file.
    ii. `isUseOriginalAlgorithm` determines whether the shuffling algorithm used is the one from the manuscript (= 1) or a newer improved algorithm that more completely randomizes the data (= 0).
3. Choose the function to use to process the selected data file (Figure 3).

```
Please select the function to start with from the selections below:

    1) Load MouseLight JSON Files and Pre-process Data
    2) Shuffle Data
    3) Statistically Analyze Data
    4) Hierarchical Clustering
    5) Analysis of Axonal Divergence
    6) Analysis of Axonal Convergence
    7) Soma Analysis
    8) NNLS Analysis
    9) Strahler Order Analysis of the Presubiculum
    !) Exit

Your selection:
```

Figure 3. Main menu selection screen.

   i. The potential functions to be called can be expanded by editing the `choices` variable in the function *select_function()*, which is found in the "lib" folder, and any additional functions will have to be added to *elseif* selections in *main_menu_selector()*.
   ii. Select 'Shuffle Data' to create a randomized version of the original data file in the manner described by Wheeler et al. (accepted).
       1. Select the data-file name base to process (Figure 4).

```
Please select the data file name to process from the selections below:

    1) PRE_matrix__axonal_counts_per_neuron_per_parcel
    !) Exit

Your selection:
```

Figure 4. Data file selection screen.

   a. All possible data name bases are stored in the file "data_file_name_bases.xlsx." The file name is stored in the variable dataFileNameBase, e.g., "PRE_matrix__axonal_counts_per_neuron_per_parcel," and the ".xlsx" suffix is automatically applied to assigned to dataFileNameBase when it is needed.

2. The function *shuffle_raw_updated()* is called automatically, where the parameters being passed are the input file name, the label for the number of shuffles, the number of shuffles, and the flag determining which shuffling algorithm to use.
3. The function *shuffle_matrix()* is called automatically, where the parameters being passed are the original data matrix, the number of shuffles, and the flag determining which shuffling algorithm to use.
    a. A value from the data matrix is selected, as is another value from a different row and a different column.
    b. For the original shuffling algorithm, the maximum possible value (`maxBound`) to be swapped is determined from the four values that are derived from the two rows and two columns involved in the swap by always starting by directly comparing the top left value to the bottom right value.
    c. For the newer improved shuffling algorithm, which balances whether the top-left to bottom-right or the top-right to bottom-left diagonal is considered first, find the minimal values comparing the top-left to the bottom-right (`minTopLeftBottomRight`) and comparing the top-right to the bottom-left (`minTopRightBottomLeft`).
    d. The randomized data matrix is automatically saved to the "data/" and "output/" directories.
        i. The auto-generated example randomized output file is named in the manner of "PRE_matrix__axonal_counts_per_neuron_per_parcel__100K__fully_shuffled_yyyymmddHHMMSS.xlsx," where 'yyyy' is the 4-digit year, 'mm' is the 2-digit month, 'dd' is the 2-digit day, 'HH' is the 2-digit hour in 24-hour format, 'MM' is the 2-digit minute, and 'SS' is the 2-digit seconds.
iii. Select 'Statistically Analyze Data' to statistically compare the original data file to the randomized data file and to generate the associated histograms, which tally the angles between the count vectors for the original data and for the randomized data.
    1. Select the data-file name base to process (Figure 3).
    2. Select the shuffled data-file name base to process (Figure 5).

```
Select shuffled data file:

Please select your .xlsx file from the selections below:

    1) PRE_matrix__axonal_counts_per_neuron_per_parcel__100K__fully_shuffled_20231005154095.xlsx
    2) PRE_matrix__axonal_counts_per_neuron_per_parcel__100K__fully_shuffled_20231005154095__angles_20231005155771.xlsx
    3) PRE_matrix__axonal_counts_per_neuron_per_parcel__100K__fully_shuffled_20231023132478.xlsx
    !) Exit

Your selection:
```

Figure 5. Shuffled data file selection screen.

3. The function *determine_vector_angles()* is automatically called, where the parameters being passed are the file name of the data (either original or randomized), the label for the number of shuffles, and a value of 0 or 1 for the flag that determines whether to use the label for the number of shuffles (1 for original data and 0 for randomized data).
  a. The necessary data file is loaded, where the original data file is loaded automatically, and the randomized data is loaded after a selection is made from a presented menu listing, such as "PRE_matrix__axonal_counts_per_neuron_per_parcel__100K__fully_shuffled_yyyymmddHHMMSS.xlsx."
  b. The function *compute_angle_between_vectors()* is automatically called, where the parameter being passed is the data matrix. The count row vectors in the data matrix are compared two at a time by taking the arccosine of the dot product of the two row vectors to determine the angle between them, and a matrix is returned, which contains the number of the first row vector being compared in the first column, the number of the second row vector being compared in the second column, and the angle between the two row vectors in the third column (Figure 6).

| 1 | 2 | 78 |
|---|---|----|
| 1 | 3 | 25 |
| 1 | 4 | 18 |
| 1 | 5 | 87 |
| 1 | 6 | 47 |

Figure 6. Vector differences data file.

  c. The data are automatically saved to the "data/" and "output/" directories. The auto-generated file for the original data is named in the manner of "PRE_matrix__axonal_counts_per_neuron_per_parcel__100K__angles_yyyymmddHHMMSS.xlsx," and the auto-generated example file for the randomized data is named in the manner of "PRE_matrix__axonal_counts_per_neuron_per_parcel__100K__fully_shuffled_yyyymmddHHMMSS__angles_yyyymmddHHMMSS.xlsx."
4. The function *create_scaled_histogram()* is automatically called, where the parameters being passed are the differences matrix for the original data, the differences matrix for the randomized data, and the label for the number of shuffles. The parameters being returned are the p-value and the statistics resulting from the one-tailed Levene's test between the original and randomized data.
  a. A figure is generated (Figure 7), which overlays the histogram of the original data (blue) with that of the randomized data (orange), and it is saved to the "output/" directory in a MATLAB figure file named in the manner of "PRE_matrix__axonal_counts_per_neuron_per_parcel__100K__histogram_yyyymmddHHMMSS.fig" and to a PNG file named in the manner of "PRE_matrix__axonal_counts_per_neuron_per_parcel__100K__histogram_yyyymmddHHMMSS.png."

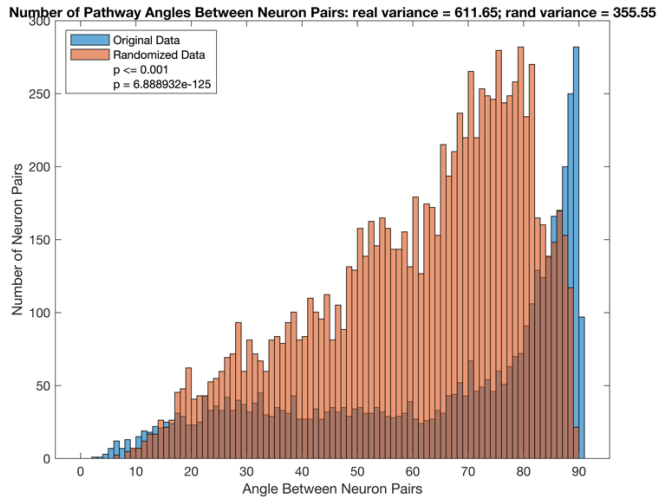
Figure 7. Histogram of angle differences between the original and shuffled data.

    b. Reported are the variances of the distributions of the original and randomized data, and the p-value resulting from a statistical comparison of the original and randomized data using a one-tailed Levene's test to determine if the variance of the original data is significantly larger than the variance of the randomized data.
  iv. Select 'Hierarchical Clustering' to generate a clustering dendrogram, where the original data are statistically compared to a randomized version of the original data by performing a one-tailed Levene's test at each potential branching point.
    1. Select the data-file name base to process (Figure 3).
    2. Select the vector differences file for the original data (Figure 8).

```
Select angles data file:

Please select your .xlsx file from the selections below:

    1) PRE_matrix__axonal_counts_per_neuron_per_parcel__100K__angles_20231005155771.xlsx
    2) PRE_matrix__axonal_counts_per_neuron_per_parcel__100K__angles_20231023132608.xlsx
    !) Exit

Your selection:
```
Figure 8. Angles data file selection screen.

    3. The function *hierarchical_clustering()* is automatically called, where the parameters being passed are the file name for the vector differences for the original data, the label for the number of shuffles, the number of shuffles, and the flag determining which shuffling algorithm to use.
      a. The dendrogram is generated using the built-in *linkages()* function with the 'average' parameter setting (Figure 9) and is saved to the "output/" directory in a MATLAB figure file named in the manner of "PRE_matrix__axonal_counts_per_neuron_per_parcel__100K__average_dendrogram.fig" and in a PNG file named in the manner of "PRE_matrix__axonal_counts_per_neuron_per_parcel__100K__average_dendrogram.png."

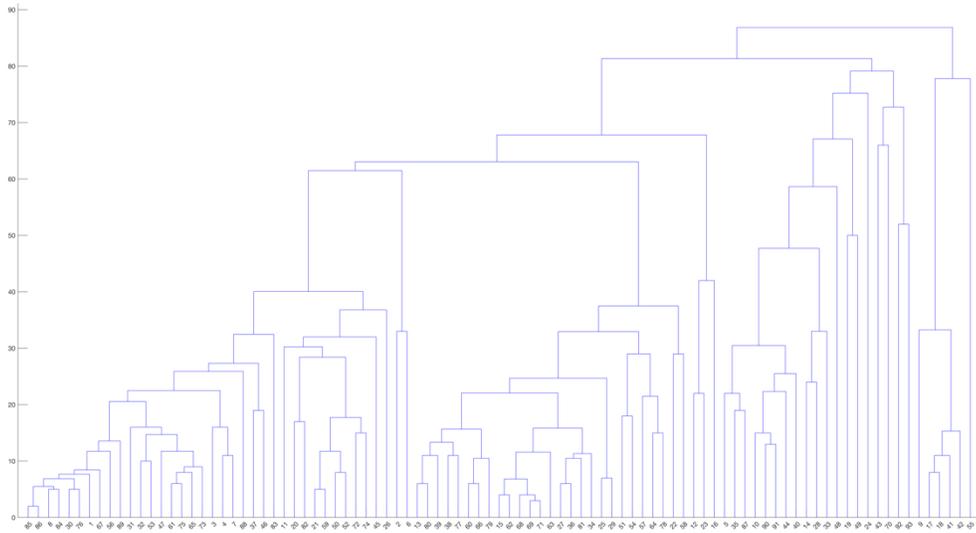

Figure 9. Hierarchical clustering dendrogram.

      b. A listing of which neuron entries (or row numbers) in the original dataset correspond to separate clusters is generated and saved to the "output/" directory with a format in the manner of "PRE_matrix__axonal_counts_per_neuron_per_parcel__neuron_numbers_by_cluster_yyyymmddHHMMSS.xlsx."

      c. A listing of which neuron names (or row labels) in the original dataset correspond to separate clusters is generated and saved in five separate files. The listing is saved to the "output/" directory and is named in the manner of "PRE_matrix__axonal_counts_per_neuron_per_parcel__neuron_names_by_cluster_yyyymmddHHMMSS.xlsx." The listing is also saved to the "data/" directory and is named in the manner of "divergence__PRE_matrix__axonal_counts_per_neuron_per_parcel__neuron_names_by_cluster_yyyymmddHHMMSS.xlsx," "convergence__PRE_matrix__axonal_counts_per_neuron_per_parcel__neuron_names_by_cluster_yyyymmddHHMMSS.xlsx," "somata__PRE_matrix__axonal_counts_per_neuron_per_parcel__neuron_names_by_cluster_yyyymmddHHMMSS.xlsx," and "nnls__PRE_matrix__axonal_counts_per_neuron_per_parcel__neuron_names_by_cluster_yyyymmddHHMMSS.xlsx."

      d. **NNLS preparation**: A listing of the counts per parcel (for neuronal data) per cluster is generated in preparation for the non-negative least squares analysis and saved to the "output/" directory with a format in the manner of "PRE_matrix__axonal_counts_per_parcel_per_cluster_yyyymmddHHMMSS.xlsx."

**Procedures for Janelia MouseLight-Specific Data.**

1. Choose the function to use to process the selected data file.
    i. Select 'Load MouseLight JSON Files and Pre-process Data' to load all the MouseLight JSON files, which describe the locations of all the axonal, dendritic, and somatic points that make up a given neuronal reconstruction and save the data to various files. The parameter being passed is the current date string, `nowDateStr`, and the parameter being returned is the matrix `morphologyMatrix` of tallies of all axonal and dendritic points and the matrix of locations of all somata. All the MouseLight JSON files are loaded from the directory "data/Mouse_Neurons/MouseLight-neurons/."
        1. The matrix of tallies of all axonal and dendritic points and the matrix of locations of all somata are saved to files in the "output/" directory.
            a. The axonal and dendritic counts are saved to a file named in the manner of "ALL__axon-dendrite_counts_yyyymmddHHMMSS.xlsx," the associated neuron names are saved to a file named in the manner of "ALL__neurons_yyyymmddHHMMSS.xlsx," the associated targeted parcels are saved to a file named in the manner of "ALL__parcels_yyyymmddHHMMSS.xlsx," the soma locations are saved to a file named in the manner of "ALL__soma_locations_yyyymmddHHMMSS.xlsx," a binary summary of which parcels are invaded by axons is saved to a file named in the manner of "ALL__axonal_parcel_invasions_yyyymmddHHMMSS.xlsx," and a summary of the axonal counts per parcel is saved to a file named in the manner of "ALL__axonal_counts_summary_yyyymmddHHMMSS.xlsx."
        2. The function *filter_matrix()* is automatically called in which the columns of the matrix, where a particular parcel has no associated axonal counts, are filtered out, and the final listing of remaining parcels and the resulting matrix of axonal counts are saved to a file, where the parameters being passed are `morphologyMatrix`, the region prefix, and `nowDateStr`, and the parameter being returned is the filtered version of `morphologyMatrix` called `filteredRawMatrix`.
            a. The filtered file of parcels is saved to the "data/" and "output/" directories and is named in the manner of "ALL__filtered_parcels_yyyymmddHHMMSS.xlsx," and the matrix of axonal counts is saved to a file in the "data/" and "output/" directories and is named in the manner of "ALL__axonal_counts_per_parcel_yyyymmddHHMMSS.xlsx."
    ii. Select 'Analysis of Axonal Divergence' to analyze the extent of divergence from the source of the clusters (e.g., the presubiculum) to all targeted parcels selected by the user. The program automatically targets both the ipsilateral and contralateral parcels from the prepared lists when computing the number of axonal point invasions in each targeted parcel.
        1. Pre-prepare lists of the parcels targeted by each cluster and save them to files in the "data/" directory in the manner of "PRE__divergence_parcels__clusterC3.xlsx," where "PRE" represents

the originating parcel abbreviation and "3" represents the number of neurons associated with the originating parcel cluster "C."
    a. The list of potential targeted parcels can be found in the "data/" directory in a pre-prepared file named something in the manner of "PRE__parcel_names.xlsx."
2. Select the divergence source file from the "data/" directory (Figure 10), stored in the variable `neuronNamesByClusterCellArray`, that groups the neuron names (or row labels) by cluster, where the file begins with the phrase "divergence" and is loaded automatically once selected, in the manner of "divergence__PRE_matrix__axonal_counts_per_neuron_per_parcel__neuron_names_by_cluster_yyyymmddHHMMSS.xlsx."

```
Please select a file of neuron names by cluster for an analysis of axonal divergence.

Please select your .xlsx file from the selections below:

    1) divergence__PRE_matrix__axonal_counts_per_neuron_per_parcel__neuron_names_by_cluster_20231005171414.xlsx
    2) divergence__PRE_matrix__axonal_counts_per_neuron_per_parcel__neuron_names_by_cluster_20231023133435.xlsx
    !) Exit

Your selection:
```
Figure 10. Divergence analysis source file selection screen.

3. Enter the abbreviation for the originating parcel for the divergence analysis, which is stored in the variable `parcelAbbreviation`.
    a. Formal parcel abbreviations can be found in the "brainAreas.json" file located in the "data/" directory, e.g., "PRE" for "presubiculum."
4. The function *analysis_divergence()* is automatically called, which automatically steps through all axonal points for each neuron in each cluster to determine the total axonal length between points in each targeted ipsilateral or contralateral parcel, where the parameters being passed are `neuronNamesByClusterCellArray` and `parcelAbbreviation`. The sets of axonal lengths for the targeted parcels are compared to each other using a Wilcoxon Signed Rank Test, and the resulting p values from the set of statistical tests are corrected for by False Discovery Rate to determine their significance.
    a. A box and whisker plot is generated using the built-in *boxplot()* function and is saved to the "output/" directory in a MATLAB figure file named in the manner of "PRE__divergence_box_and_whisker_plot__for_clusterE6_yyyymmddHHMMSS.fig" and in a PNG file named in the manner of "PRE__divergence_box_and_whisker_plot__for_clusterE6_yyyymmddHHMMSS.png."
    b. The results of all statistical tests are saved to files in the "output/" directory in files named in the manner of "PRE__clusterE6__Wilcoxon_and_FDR_yyyymmddHHMMSS.xlsx."

iii. Select 'Analysis of Axonal Convergence' to analyze the extent of convergence from all the clusters in an originating parcel (e.g., the presubiculum) to each targeted parcel selected by the user.
1. Pre-prepare a list of the parcels targeted by each cluster and save them to files in the "data/" directory in the manner of "PRE__convergence_parcels.xlsx."
2. Select the convergence source file from the "data/" directory (Figure 11), stored in the variable `neuronNamesByClusterCellArray`, that groups the neuron names (or row labels) by cluster, where the file begins with the phrase "convergence" and is loaded automatically once selected, in the manner of "convergence__PRE_matrix__axonal_counts_per_neuron_per_parcel__neuron_names_by_cluster_yyyymmddHHMMSS.xlsx."

```
Please select a file of neuron names by cluster for an analysis of axonal convergence.

Please select your .xlsx file from the selections below:

    1) convergence__PRE_matrix__axonal_counts_per_neuron_per_parcel__neuron_names_by_cluster_20231005171414.xlsx
    2) convergence__PRE_matrix__axonal_counts_per_neuron_per_parcel__neuron_names_by_cluster_20231023133435.xlsx
    !) Exit

Your selection:
```
Figure 11. Convergence analysis source file selection screen.

3. Enter the abbreviation for the originating parcel for the convergence analysis, which is stored in the variable `parcelAbbreviation`.
4. The program automatically loads the convergence parcels into the variable `parcelsCellArray` from a file named in the manner of "PRE__convergence_parcels.xlsx."
5. The function *analysis_convergence()* is automatically called, which automatically steps through all axonal points for each neuron in each ipsilateral or contralateral half of the cluster to determine the total axonal length between points in each targeted parcel, where the parameters being passed are `neuronNamesByClusterCellArray`, `parcelsCellArray`, and `parcelAbbreviation`. The sets of axonal lengths for the targeted parcels are compared to each other using a Wilcoxon Signed Rank Test, and the resulting p values from the set of statistical tests are corrected for by False Discovery Rate to determine their significance.
    a. Box and whisker plots are generated using the built-in *boxplot()* function and are saved to the "output/" directory in MATLAB figure files named in the manner of "PRE__convergence_box_and_whisker_plot__(I)Entorhinal area, lateral part_ yyyymmddHHMMSS.fig" and in a PNG file named in the manner of "PRE__convergence_box_and_whisker_plot__(I)Entorhinal area, lateral part_ yyyymmddHHMMSS.png."
    b. The results of all statistical tests are saved to files in the "output/" directory in files named in the manner of

"PRE__(l)Entorhinal area, lateral part__Wilcoxon_and_FDR_yyyymmddHHMMSS.xlsx."
iv. Select 'Soma Analysis' to analyze the physical distribution of somata locations (e.g., in the presubiculum) via the determination of convex hulls for the associated clusters.
1. Select the somata analysis source file from the "data/" directory (Figure 12), which groups the neuron names (or row labels) by cluster, where the file begins with the phrase "somata," is loaded automatically once selected into the variable `neuronNamesByClusterCellArray` and is named in the manner of "somata__PRE_matrix__axonal_counts_per_neuron_per_parcel__neuron_names_by_cluster_yyyymmddHHMMSS.xlsx."

```
Please select a file of neuron names by cluster for a somatic convex hull analysis.

Please select your .xlsx file from the selections below:

    1) somata__PRE_matrix__axonal_counts_per_neuron_per_parcel__neuron_names_by_cluster_20231005171414.xlsx
    2) somata__PRE_matrix__axonal_counts_per_neuron_per_parcel__neuron_names_by_cluster_20231023133435.xlsx
    !) Exit

Your selection:
```

Figure 12. Somatic analysis data file selection screen.

2. Enter the abbreviation for the originating parcel for the soma analysis.
3. The function *convex_hull_outliers()* is automatically called, which determines the somata locations for the two clusters being analyzed, infers which somata are co-localized in the two clusters, and computes the volume fraction of the co-localization based on the convex hull volumes of the two clusters. The parameters being passed are `neuronNamesByClusterCellArray` and `parcelAbbreviationStr`.
4. Select the first cluster to be analyzed.
5. Select the second cluster to be analyzed.
6. Multiple output files are generated and saved to the "/output" directory.
   a. A figure is generated of the convex hull volumes of the two clusters and is saved to a MATLAB figure file named in the manner of "PRE__clusterD19_and_clusterA38__convex_hulls_with_outliers_yyyymmddHHMMSS.fig" and to a PNG file named in the manner of "PRE__clusterD19_and_clusterA38__convex_hulls_with_outliers_yyyymmddHHMMSS.png."
   b. A figure is generated of the somata locations of the two clusters and is saved to a MATLAB figure file named in the manner of "PRE__clusterD19_and_clusterA38__somata_locations_yyyymmddHHMMSS.fig" and to a PNG file named in the manner of "PRE__clusterD19_and_clusterA38__somata_locations_yyyymmddHHMMSS.png."
   c. A figure is generated of the co-localized somata locations of the two clusters overlayed with the convex hull volume of the first cluster and the somata locations of the second cluster and is saved to a MATLAB figure file named in the manner of "PRE__clusterD19_somata_locations_in_intersection_with_clusterA38_yyyymmddHHMMSS.fig" and to a PNG file named in the manner of "PRE__clusterD19_somata_locations_in_intersection_with_clusterA38_yyyymmddHHMMSS.png."

d. A figure is generated of the co-localized somata locations of the two clusters overlayed with the convex hull volume of the second cluster and the somata locations of the first cluster and is saved to a MATLAB figure file named in the manner of "PRE__clusterA38_somata_locations_in_intersection_with_clusterD19_yyyymmddHHMMSS.fig" and to a PNG file named in the manner of "PRE__clusterA38_somata_locations_in_intersection_with_clusterD19_yyyymmddHHMMSS.png."

e. A figure is generated of the co-localized somata locations of the two clusters overlayed with the convex hull volumes of the two clusters and is saved to a MATLAB figure file named in the manner of "PRE__somata_locations_in_intersection_of_clusterD19_and_clusterA38_yyyymmddHHMMSS.fig" and to a PNG file named in the manner of "PRE__somata_locations_in_intersection_of_clusterD19_and_clusterA38_yyyymmddHHMMSS.png."

f. A text file containing the overlap volume percentage of the co-localized somata locations of the two clusters is saved to a file named in the manner of "PRE__overlap_volume_of_clusterD19_and_clusterA38_is_15_percent_yyyymmddHHMMSS.txt."

v. Select 'NNLS Analysis' to perform a non-negative least squares analysis.

1. Pre-prepare the data for this analysis by taking the data from the step **NNLS preparation** and adding a last column of tract values from a source, such as the regional connectivity from anterograde tracing to the m known targets, as presented in the Allen Mouse Brain Connectivity Atlas (http://connectivity.brain-map.org/projection).

2. Select the pre-prepared file (Figure 13), which is named in the manner of "nnls__PRE_matrix__axonal_counts_per_parcel_per_cluster_and_tract_values.xlsx."

```
Please select a file of axonal counts per parcel per cluster for non-negative least squares analysis.

Please select your .xlsx file from the selections below:

    1) nnls__PRE_matrix__axonal_counts_per_neuron_per_parcel__neuron_names_by_cluster_20231005171414.xlsx
    2) nnls__PRE_matrix__axonal_counts_per_neuron_per_parcel__neuron_names_by_cluster_20231023133435.xlsx
    3) nnls__PRE_matrix__axonal_counts_per_parcel_per_cluster_and_tract_values.xlsx
    !) Exit

Your selection:
```

Figure 13. Non-negative least squares analysis data file selection screen.

3. The function *nnls()* is called, which automatically bi-normalizes the data and applies non-negative least square to the data, where the parameters being passed are `axonalCountsPerParcelPerClusterCellArray` and the name of the pre-prepared file. Multiple output files are generated and saved to the "/output" directory.
    a. Data are normalized by row and saved to a file named in the manner of "nnls__PRE_matrix__axonal_counts_per_parcel_per_cluster_and_tract_values__row_normalized_yyyymmddHHMMSS.xlsx."
    b. Data are scaled, where each value is divided by the number of regions and multiplied by the number of clusters, such that the sum of all values is equal to the number of clusters and saved to a file named in the manner of "nnls__PRE_matrix__axonal_counts_per_parcel_per_cluster_and_tract_values__scaled_yyyymmddHHMMSS.xlsx."
    c. Data are normalized by column and saved to a file named in the manner of "nnls__PRE_matrix__axonal_counts_per_parcel_per_cluster_and_tract_values__column_normalized_yyyymmddHHMMSS.xlsx."
    d. Tract values are normalized, and the final bi-normalization data are saved to a file named in the manner of "nnls__PRE_matrix__axonal_counts_per_parcel_per_cluster_and_tract_values__bi_normalized_yyyymmddHHMMSS.xlsx."
    e. The resulting vector x, where x is the k-dimensional vector representing the fractions of neurons in each neuronal class, is saved in a file named in the manner of "nnls__PRE_matrix__axonal_counts_per_parcel_per_cluster_and_tract_values__X_vector_yyyymmddHHMMSS.xlsx."
    f. The squared Euclidean norm of the residual of the MATLAB function *lsqnonneg()* is calculated and the result saved in a file named in the manner of "nnls__PRE_matrix__axonal_counts_per_parcel_per_cluster_and_tract_values__residual_norm_yyyymmddHHMMSS.xlsx."
vi. Select 'Strahler Order Analysis of the Presubiculum' to calculate the Strahler order values of the axonal branches in the presubiculum subset of the MouseLight dataset. The function *Strahler()* is automatically called with no parameters, which loads presubiculum-related SWC files and generates axonal branch-related statistics. Data are stored the "data/Mouse_Neurons/MouseLight_PRE-SWC_files/" directory in files named in the manner of "AA0021.swc." Multiple output files are generated and saved to the "/output" directory.
    1. On a per neuron basis, the mean branch length, the number of branches, and the mean number of reconstruction nodes are listed as a function of Strahler order number and are saved to the "/output" directory in a file named in the manner of "branch_statistics_per_neuron__yyyymmddHHMMSS.xlsx."

2. On a per branch basis, the branch length, number of reconstruction nodes per branch, and the Strahler order number are listed and are saved to the "/output" directory in a file named in the manner of "branch_statistics_per_branch__yyyymmddHHMMSS.xlsx."
   3. New copies of the input SWC files are generated, where reconstruction node type is changed from a value of 2 to a value of 5 when the node's Strahler order number is in the range of 1-3 and are saved in a sub-directory of the "/output" directory named in the manner of "MouseLight_PRE_SWC_files_modified_yyyymmddHHMMSS" in files named in the manner of "AA0021_modified.swc."

**Conclusions.**

Presented are a pair of protocols for the analysis of cellular data: one that involves a general approach to cellular classification and another that explores specific characteristics of neuronal data from the Janelia MouseLight project. Cellular classification is achieved using unsupervised hierarchical clustering that has been enhanced by an associated statistical test. The described computer code is freely available[22] from a GitHub repository (https://github.com/Projectomics/MATLAB) to facilitate its adoption for the study of wide varieties of cellular data.

**Acknowledgements.**

This work was supported in part by NIH grants R01NS39600, U01MH114829, and RF1MH128693. This methodology was used in Wheeler et al. "Unsupervised classification of brain-wide axons reveals neuronal projection blueprint: an illustrative application to the presubiculum"[16].

**Competing Interests.**

The authors declare no competing financial interests.

**Ethics.**

All original data analyzed using this protocol were published previously in accordance with the authors' respective ethics committees.

**References.**